\documentclass[aps,floatfix,nofootinbib,twocolumn,prl]{revtex4-1}

\usepackage{graphicx}
\usepackage{epic}
\usepackage{eepic}
\usepackage{latexsym}

\newcommand{\eq}[1]{(\ref{#1})}
\newcommand{\be}{\begin{equation}}
\newcommand{\ee}{\end{equation}}
\newcommand{\bea}{\begin{eqnarray}}
\newcommand{\eea}{\end{eqnarray}}

\newcommand{\vs}[1]{\vspace{#1 mm}}
\newcommand{\hs}[1]{\hspace{#1 mm}}

\def\fr{\frac}

\def\l{\lambda}

\def\m{\mu}
\def\n{\nu}

\def\r{\rho}

\def\del{\partial}

\let\bm=\bibitem
\def\nn{\nonumber}

\begin{document}
\large

\title{ \Large Hubble's law and faster than light expansion speeds}

\author{Ali Kaya}
\email[]{ali.kaya@boun.edu.tr}
\affiliation{\large Bo\~{g}azi\c{c}i University, Department of Physics, \\ 34342,
Bebek, Istanbul, Turkey }

\begin{abstract}

Naively applying Hubble's law to a sufficiently distant object gives a receding velocity larger than the speed of light.  By discussing a very similar situation in special relativity, we argue that Hubble's law is meaningful only  for nearby objects with non-relativistic receding speeds.  To support  this claim, we note that in a curved spacetime manifold it is not possible to directly compare tangent vectors at different points, and thus there is no natural definition of relative velocity between two spatially separated objects in cosmology. We clarify the geometrical meaning of the Hubble's receding speed $v$ by showing that in a Friedmann-Robertson-Walker spacetime if  the four-velocity vector of  a comoving object is parallel-transported along the straight line in flat comoving coordinates to the position of a second comoving object, then $v/c$ actually becomes the {\it rapidity} of the local Lorentz transformation, which maps the fixed four-velocity vector to the transported one. 

\end{abstract}

\maketitle

Hubble's observations, which showed that the distant galaxies recede with speeds proportional to their distances, marked the beginning of a new era for cosmology \cite{hub}. Hubble's results not only changed the widely accepted steady state universe idea, but also provide evidence for the general theory of relativity, which treats spacetime as a dynamical object. Indeed, before Hubble published his observations Friedmann had showed that the Einstein's equations imply an expanding universe \cite{fr}. Therefore, in courses about  cosmology or general relativity one should certainly discuss Hubble's law and its consequences. 

It is very easy to derive Hubble's law in the context of a (flat) Friedmann-Robertson-Walker  cosmology, which has the line element
\bea\label{frw} 
ds^2=-c^2dt^2+a(t)^2\left(dx^2+dy^2+dz^2\right). 
\eea
The scale factor of the universe $a(t)$ can be thought to define physical lengths so that multiplying  the {\it coordinate} distance between two points in the {\it space} by $a(t)$ gives the {\it physical} distance between these points at time $t$.  Objects which have constant spatial coordinates $(x,y,z)$ are said to be {\it comoving}.\footnote{The worldline of a comoving object can be parametrized as $t=\l$, $x=x_0$, $y=y_0$ and $z=z_0$. Using this parametrization and \eq{chr}, it can easily be shown that  the worldline is a geodesics of \eq{frw}, i.e. $d^2x^\m/d\l^2+\Gamma^\m{}_{\n\r}(dx^\n/d\l)(dx^\r/d\l)=0$.} The coordinate (or comoving) distance $D_c$ between  two comoving objects with  coordinates $(x_1,y_1,z_1)$ and $(x_2,y_2,z_2)$ is given by $D_c=\sqrt{(x_1-x_2)^2+(y_1-y_2)^2+(z_1-z_2)^2}$. The comoving distance $D_c$ does not change with time since comoving objects are not moving in space and have  fixed positions. However,  the physical spatial distance $D$  becomes $D=a(t) D_c$. Differentiating $D$ with respect to time $t$ gives the Hubble's law
\be\label{h}
v=HD,
\ee
where  the time derivative of the distance is interpreted as the speed,  $v=\dot{D}$, and the Hubble parameter is defined as
\be
H=\fr{\dot{a}}{a},
\ee 
where the dot denotes the time derivative. 

Despite the simplicity of its derivation, Hubble's law raises an immediate concern for students: what about the very distant objects whose receding speeds exceed  the speed of light? In other words, what happens to the relativistic assumption that nothing can move faster than light? The usual answer to such questions is that the receding speed does not correspond to a physical propagation because the two objects considered in the above derivation are not moving at all.  Thus there is no problem for $v$ in \eq{h} to be larger than the speed of light. 

Although this explanation of Hubble's law is certainly correct, one may  encounter  deeper questions.  If the receding velocity due to the expansion of the universe does not quantify a real propagation, what does it  physically correspond? Isn't it the case that to determine the Hubble's constant $H$, astronomers try to  measure the receding velocities  of galaxies? How can the measured velocity of an object can be different than the physical propagation velocity of that object with respect to the observer, and if not, how  can the propagation velocity be larger than $c$? 

There are many misleading statements in the literature about the superluminal receding velocities. These statements are nicely analyzed in \cite{l1,l2}. Our aim here is not to examine and criticize these statements but to point out that Hubble's law can be confusing if it is not carefully analyzed. For example, in \cite{mc} the velocities of recession greater than that of light is said to contradict  one of the basic postulates of relativity. Similarly, in the book \cite{sc} it is claimed that \eq{h} cannot be exact for large distances giving faster than light receding speeds, although we see that the above derivation is straightforward and valid for any distance. (In the same chapter in \cite{sc}, the claim is clarified by deriving luminosity distance vs. redshift relation and then showing that it reduces to Hubble's law for nearby objects.) There are also discussions in the literature about the observability of the galaxies with superluminal receding speeds (see e.g. \cite{sol1,sol2,sol3}). 

One reason for the confusion is that neither the speed $v$ nor the distance $D$ can  be directly  observed. Instead, astronomers measure redshift and luminosity (or angular diameter) distance, which are indirectly related to $v$ and $D$ (see e.g. \cite{ha} for Hubble's law expressed as a redshift-distance relation). Of course in a general relativistic framework  there must not arise any issue with the faster than light propagation. In any case, however, the relation \eq{h} needs further clarification, at least for students.  

\begin{figure}
\centerline{
\includegraphics[width=8cm]{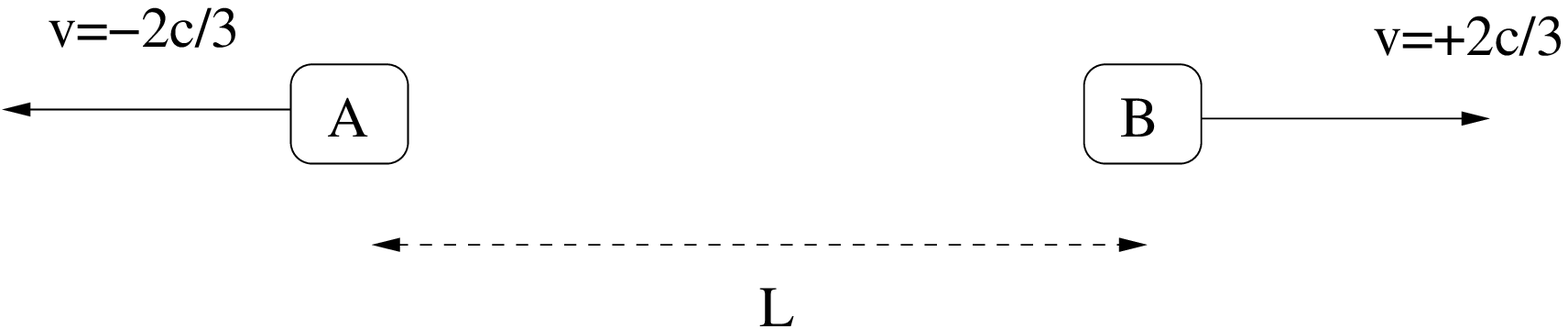}}
\vs{11}
\centerline{
\includegraphics[width=7cm]{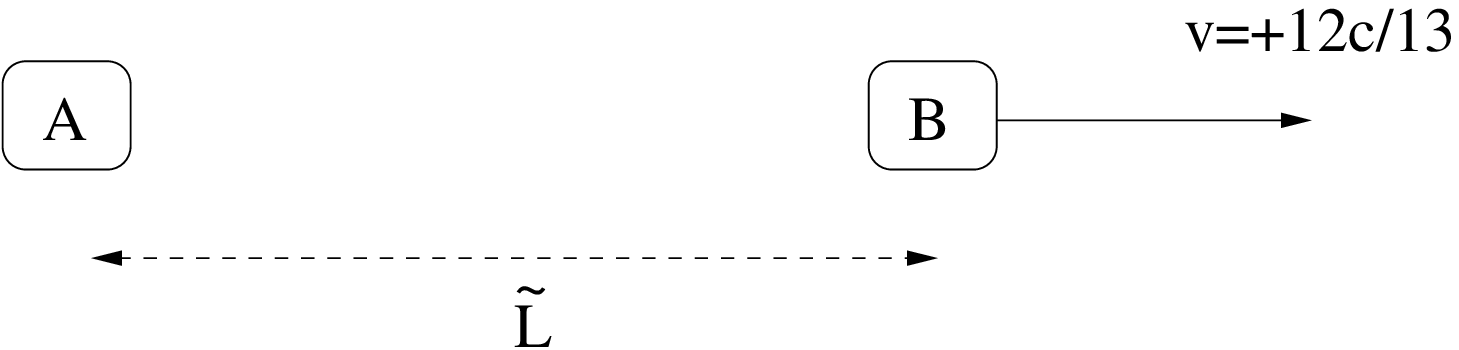}}
\caption{In an inertial frame $S$ the rate of change of the distance $L$ between the objects A and B is $4c/3$, which is larger than the speed of light. On the other hand,  in the rest frame $\tilde{S}$ of A the rate of change of the  new distance  $\tilde{L}$ with respect to the new time parameter is less than $c$} 
\label{fig1}
\end{figure}

The proper language of general relativity is differential geometry, which must also be used in discussing problems in cosmology. Below, we will analyze Hubble's law in that context. However, students who are acquainted with special relativity can also  solve the issue of superluminal receding velocities easily. Consider the following example in special relativity (see e.g. problem 3.5 in \cite{ew}). In an inertial frame $S$ two objects A and B are moving along $x$-direction with respective velocities $-2c/3$ and $+2c/3$ (see figure \ref{fig1}). In that frame, the physical distance $L$ between A and B is a well defined quantity. Actually $L$ exactly plays the role of $D$ in  Hubble's law; namely both refer to physical distances in prescribed frames. But, the  derivative of $L$ with respect to the time coordinate of the inertial frame $S$ can be found as $\dot{L}=4c/3$, which is larger than $c$. This result does not lead one to think that one of the basic postulates of relativity is violated, because $\dot{L}$ does not give the relative velocity between A and B. It is just the time derivative of a distance and has no {\it direct  physical meaning}. Only when  the speeds of A and B become much smaller than the speed of light,   $\dot{L}$   approximates the relative speed, which is  equivalent to non-relativistic velocity addition formula. To find the correct relative velocity, one should use relativistic velocity addition formula or equivalently apply a Lorentz transformation to the rest frame $\tilde{S}$ of, say A. Only then  can the  derivative of the  distance  $\tilde{L}$ in the frame $\tilde{S}$ with respect to the  time parameter of  $\tilde{S}$  be interpreted as the relative speed,  which is  $12c/13<c$ using the relativistic velocity addition formula (see figure \ref{fig1}). 

To have a better understanding of this example, let  us parametrize the position four-vectors of A and B in the frame $S$ as $x^\m_A=(ct,-2ct/3,0,0)$ and $x^\m_B=(ct,+2ct/3,0,0)$, where $t$ is the time parameter\footnote{Note that one can actually choose infinitely many different parametrizations for a position four-vector and correspondingly get different  covariant four-velocity vectors. However, the standard velocity of an object in a frame is defined with respect to the time parameter of that frame, so the above parametrization is a convenient choice.}  of $S$. The physical distance $L$ in the frame $S$ can be given as $L^2=\eta_{\m\n} (x_A^\m-x_B^\m)(x_A^\n-x_B^\n)$, where the metric is given by $\eta_{\m\n}=\textrm{diag}(-1,1,1,1)$. Although  $L$ can be recast to an invariant quantity,  its derivative with respect to time $t$ cannot. The lesson from special relativity is that  the time derivative of the distance between two objects does not necessarily give the relative speed, or to be more precise it does not correspond to a physically meaningful quantity. 

The situation with the Hubble's law $v=HD$ is  similar. If $v$ is small,  it can be approximately interpreted as the relative velocity between two comoving objects (more on this below). Otherwise, it does not have a direct physical meaning, i.e. it does not give us any physical information. Therefore, there is no need to bother about superluminal receding velocities, which would be similar to worrying about $\dot{L}>c$ in the previous example. This fact must clearly be emphasized in teaching Hubble's law to avoid confusion. 

Before discussing the issue in the language of differential geometry, let us try to clarify a possible concern that might occur about the analogy we have discussed. Although students might be convinced by our example from special relativity that the time derivative of  a distance does not necessarily give the velocity, it does correctly give the relative velocity in the rest frame of A as shown in figure \ref{fig1}. Then, how can we make sure that the frame in which the Hubble's law is expressed is not special? Can this frame, in which the comoving objects do not move, be similar to the rest frame of A?. Unfortunately  it is not possible to answer such a question without properly discussing the problem in general relativity. There is in principle no difference between the frames $S$ and $\tilde{S}$ in special relativity, since both of them are inertial. Similarly, all frames in general relativity are  on an equal footing. In our example, the definition of relative velocity turns out to be equivalent to the time derivative of the distance between two objects in a frame in which one object is stationary. 

We now discuss how Hubble's law should be understood in the general relativity. We first  emphasize the following well known but occasionally  forgotten fact from differential geometry: in a manifold  there is no natural way of comparing tangent vectors at {\it different} points unless a connection is introduced as an extra structure (see e.g. \cite{w}). One cannot  move vectors around, as is usually done in the flat  Euclidean space, where there exists a well defined global notion of what it means to be parallel. Only vectors defined in the {\it same} tangent space about a point can be compared, added or subtracted. Once a connection (or a covariant derivative associated with a metric) is introduced, parallel-transportation of a vector along a given path can be defined. In that case  a vector can be carried out by parallel-transportation along a specified curve  from its original position to the position of another vector, and then these two vectors placed in the same position can be compared with each other. However, parallel-transportation is path dependent and thus this comparison is not unique. 

These comments are valid for cosmology, despite the fact that at a fixed time the space is the usual euclidean 3-space in \eq{frw}. Therefore, there is no natural way of comparing four-velocity vectors of two spatially separated objects. Relative velocity can only be defined for two objects passing through the {\it same spacetime point}. Note that the situation is different in special relativity, where the underlying Minkowski spacetime  is flat and there is a global notion of parallel-ness. 

\begin{figure}
\centerline{
\includegraphics[width=8cm]{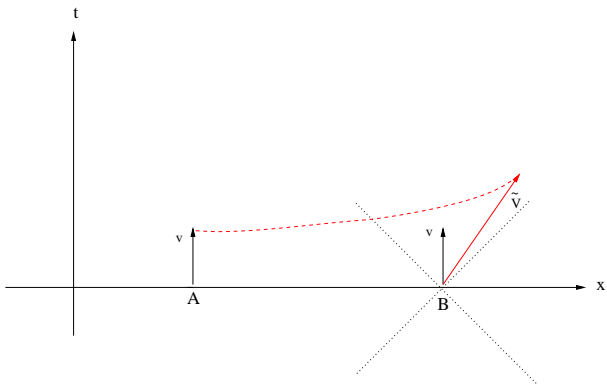}}
\caption{The parallel-transportation of the four-velocity vector of  A along the straight line joining A to  B. The dashed lines on B represent the light cone. Since parallel-transportation preserves the length of a vector, the transported vector at B is also time-like.} 
\label{fig2}
\end{figure}

Consider two comoving objects A and B in \eq{frw}, which are located along the $x$-axis with $D_c$ being the coordinate distance between them, as shown in figure \ref{fig2}. The four-velocity vectors of A and B are given by $v_A^\m=(1,0,0,0)$ and $v_B^\m=(1,0,0,0)$, where the indices refer to the coordinate basis in \eq{frw} with $x^\m=(t,x,y,z)$ (in the following discussion we use geometrical units such that $c=1$). To compare velocities let us parallel-transport $v_A^\m$ along the $x$-axis at a fixed time to the position of the object B. The tangent vector to this curve, which is  the $x$-axis, is given by
\be\label{k}
k^\m=(0,1,0,0). 
\ee
As mentioned, choosing different paths would give different vectors at B, and there is no natural definition of relative velocity. Nevertheless, we will show  that using  parallel-transportation along the $x$-axis gives a reasonable generalization of the familiar concept of relative velocity. 

The parallel-transport of a vector $v^\m$ along a curve with tangent $k^\m$ is defined as
\be\label{p}
k^\m\nabla_\m v^\n=0.
\ee
This gives a set of coupled first-order differential equation  for the components $v^\m$, which are uniquely determined once an initial vector is specified. In our example, the initial vector is $v_A^\m$ and one can see that  $v^y=v^z=0$ is the unique solution of \eq{p} with the prescribed initial conditions. For the other two components $v^t$ and $v^x$, the definition \eq{p} explicitly becomes
\bea
\del_x v^t+\Gamma^{t}{}_{xx}v^x=0,\nn\\
\del_x v^x+\Gamma^{x}{}_{xt}v^t=0,\label{pc}
\eea
where $\eq{k}$ has been used. The relevant components of the Christoffel symbol of \eq{frw} can be found as
\be\label{chr}
\Gamma^{t}{}_{xx}=a\dot{a},\hs{5}\Gamma^{x}{}_{xt}=\fr{\dot{a}}{a}.
\ee  
Defining the orthonormal frame components as 
\bea
&&v^{\hat{t}}=v^t, \nn\\
&&v^{\hat{x}}=a\, v^x,\nn
\eea
the  equation system \eq{pc} becomes 
\bea
\del_x v^{\hat{t}}+\dot{a}v^{\hat{x}}=0,\nn\\
\del_x v^{\hat{x}}+\dot{a}v^{\hat{t}}=0.\nn
\eea 
These coupled equations can be solved giving
\bea
\left[ \begin{array}{c} v^{\hat{t}}  \\ v^{\hat{x}} \end{array}\right]=\exp\left(-\left[\begin{array}{cc}0 & \dot{a}\\ \dot{a} & 0\end{array}\right] x \right) \left[ \begin{array}{c} v^{\hat{t}}_0 \\ v^{\hat{x}}_0 \end{array}\right], 
\eea
where $v^{\hat{t}}_0$ and  $v^{\hat{x}}_0$ are the components of the initial vector. The exponential can  be calculated by diagonalization, which gives
\bea\nn
\left[ \begin{array}{c} v^{\hat{t}}  \\ v^{\hat{x}} \end{array}\right]=\left[\begin{array}{cc} \cosh(\dot{a}x) & -\sinh(\dot{a}x)\\ -\sinh(\dot{a}x) &  \cosh(\dot{a}x)\end{array}\right]  \left[ \begin{array}{c} v^{\hat{t}}_0 \\ v^{\hat{x}}_0 \end{array}\right]. 
\eea
Using $x=D_c$, $D=aD_c$ and 
\be
v=HD=\dot{a}D_c,
\ee
one can determine $\tilde{v}^\m$, the transported four-velocity vector of A to the position of B, as 
\bea
\left[ \begin{array}{c} \tilde{v}^{\hat{t}}  \\ \tilde{v}^{\hat{x}} \end{array}\right]&=&\left[\begin{array}{cc} \cosh(v) & -\sinh(v)\\ -\sinh(v) &  \cosh(v)\end{array}\right]  \left[ \begin{array}{c} 1  \\ 0 \end{array}\right] \label{l}\\
&= &
\left[ \begin{array}{c} \cosh(v)  \\  -\sinh(v) \end{array}\right].\nn
\eea
Because  parallel-transportation preserves the norm, the transported vector $\tilde{v}^\m$ becomes a time-like vector obeying $\tilde{v}^\m\tilde{v}_\m=-1$ (see figure \ref{fig2}). 

We have now two vectors $v_B^\m$ and $\tilde{v}^\m$ in the same tangent space at the point B. To determine the relative speed between  $v_B^\m$ and $\tilde{v}^\m$, we note that \eq{l} is equivalent to a Lorentz transformation in the tangent space, which actually maps $v_B^\m$ to $\tilde{v}^\m$, where $v$ is the corresponding rapidity parameter related to the Lorentz factor as $\gamma=\cosh(v)$. Therefore, the  velocity giving the boost in \eq{l} can be identified as the relative velocity $v_{rel}$. Introducing the factors of $c$ and nothing that  $\gamma=1/\sqrt{1-v_{rel}^2/c^2}$, one finds 
\be\label{rel}
v_{rel}=c\,\tanh\left(\fr{v}{c}\right).
\ee
Thus, we have shown that Hubble's speed $v$ actually becomes the rapidity of the local Lorentz  transformation, which maps the original four-velocity vector of the object B to the parallel-transported four-velocity vector of the object A. In this way one sees that there is no faster than light propagation issue and from \eq{rel} the path dependent relative velocity becomes 
\be
v_{rel}<c
\ee
since $\tanh$ function is strictly less than one. 

Note that for $v\ll c$ one finds
\be\label{ri}
v_{rel} \simeq  v.
\ee
As expected, Hubble's receding velocity can be thought to give the relative velocity between two close comoving objects. It is also possible to reinterpret \eq{ri} by reversing the logic. Since Hubble's receding velocity is supposed  to give the relative velocity, it is natural to define $v_{rel}$ by parallel-transportation along the straight line joining two comoving objects at a fixed time, at least for nearby objects. For larger distances, one can imagine infinitesimally separated comoving observers placed in between A and B. Nearby observers can measure their relative velocities at a fixed time. From that information, relative velocity between distant objects A and B can be determined by integration, which indeed corresponds to parallel-transportation along the finite line-segment and thus we obtain \eq{rel}.  Consequently, $v_{rel}$ can be seen to generalize the usual concept of relative velocity in a cosmological context. 

In summary, we believe that the best way to resolve concerns about superluminal expansion speeds is to emphasize that Hubble's law does not  make sense for large distances.  We showed that if  the time derivative of the distance between two objects is naively identified as the relative velocity, then faster than light speeds can also be found in special relativity. Therefore, we need to be  careful in determining the correct physical meaning of a mathematical quantity in a relativistic theory, which is also the main issue with Hubble's law. These examples can be used to convince students that there is nothing wrong with a naive superluminal expansion speed since it has nothing to do with relative velocity or as a matter of fact  it has no direct physical significance. Moreover, we pinned down  the correct differential geometrical meaning of the Hubble's receding velocity as the rapidity of a local Lorentz transformation. With the derivation of this last result, there must not arise any further issue with faster than light expansion speeds.

\end{document}